# Technical Considerations when using Verasonics Research Ultrasound Platform for Developing a Photoacoustic Imaging System

KARL KRATKIEWICZ,[1,¥] RAYYAN MANWAR,[1,2,¥] YANG ZHOU,[1] MOEIN MOZAFFARZADEH,[3] AND KAMRAN AVANAKI,[1,2,3,*]

# Technical Considerations when using Verasonics Research Ultrasound Platform for Developing a Photoacoustic Imaging System

KARL KRATKIEWICZ,[1,¥] RAYYAN MANWAR,[1,2,¥] YANG ZHOU,[1] MOEIN MOZAFFARZADEH,[3] AND KAMRAN AVANAKI,[1,2,3,*]

[1]Wayne State University, Department of Biomedical Engineering, Detroit, USA
[2] University of Illinois at Chicago, Department of Bioengineering, Chicago, USA
[3] Tarbiat Modares University, Department of Biomedical Engineering, Tehran, Iran
[4] University of Illinois at Chicago, Department of Dermatology, Chicago, USA
[¥]These authors have contributed equally
*avanaki@uic.edu

**Abstract:** Photoacoustic imaging (PAI) is an emerging functional and molecular imaging technology that has attracted much attention in the past decade. Recently, many researchers have used the Vantage system from Verasonics® for simultaneous ultrasound (US) and photoacoustic (PA) imaging. This was the motivation to write on the details of US/PA imaging system implementation and characterization using Verasonics platform. We have discussed the experimental considerations for linear array based PAI due to its popularity, simple setup, and high potential for clinical translatability. Specifically, we describe the strategies of US/PA imaging system setup, signal generation, amplification, data processing and study the system performance.

## 1. Introduction

Photoacoustic imaging (PAI), also called optoacoustic imaging, is a three-dimensional (3-D) imaging modality that works based on the photoacoustic (PA) effect [1]. The sample (light absorbent) to be imaged is optically excited, leading to a transient temperature rise, resulting in a thermoelastic expansion of the absorber followed by emission of acoustic waves. The absorber could be endogenous such as hemoglobin (oxy- or deoxy-), myoglobin (oxy- or deoxy-), melanin, lipid, Bilirubin, water, or an exogenous contrast agent such as dyes [2]. The absorption spectrum of some of the endogenous and exogenous absorbers are shown in Figure 1; the absorption spectrum of two example imaging targets that are commonly used for calibration and performance evaluation in PAI experiments, i.e., vinyl black tape and carbon pencil lead, are given in Appendix A.

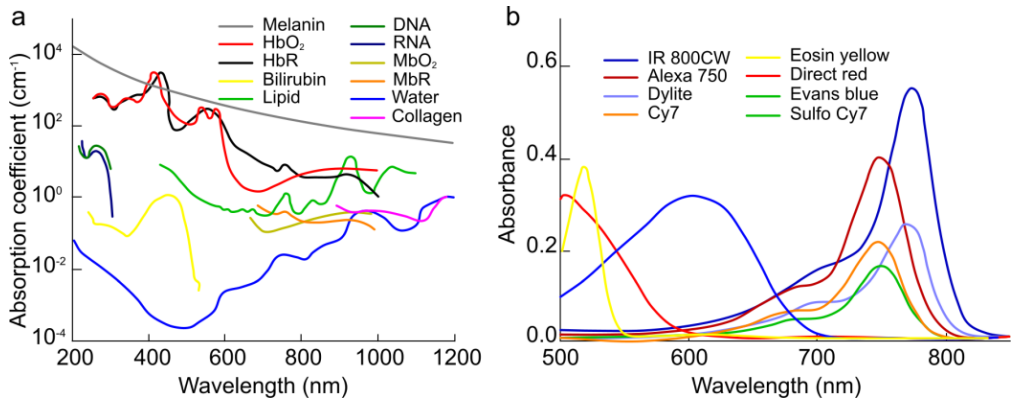

Fig. 1. Absorption spectrum of **(a)** endogenous, and **(b)** exogenous contrast agents. Reproduced with permission from [3-5]. According to Beer-Lambert law, absorbance is defined by $A = \varepsilon(\lambda) C d$ where, $\varepsilon(\lambda)$, $C$, and $d$ are molar absorptivity, concentration, and cross-section thickness of the contrast agent, respectively. Absorption coefficient is expressed as: $\alpha = (A/d) \log_{10} e$ [6].

The emitted acoustic waves from the absorber are detected by ultrasound (US) transducers. The transducer signals are then given to an image reconstruction algorithm to generate the absorption map of the tissue. The PAI process steps are illustrated in Figure 2.

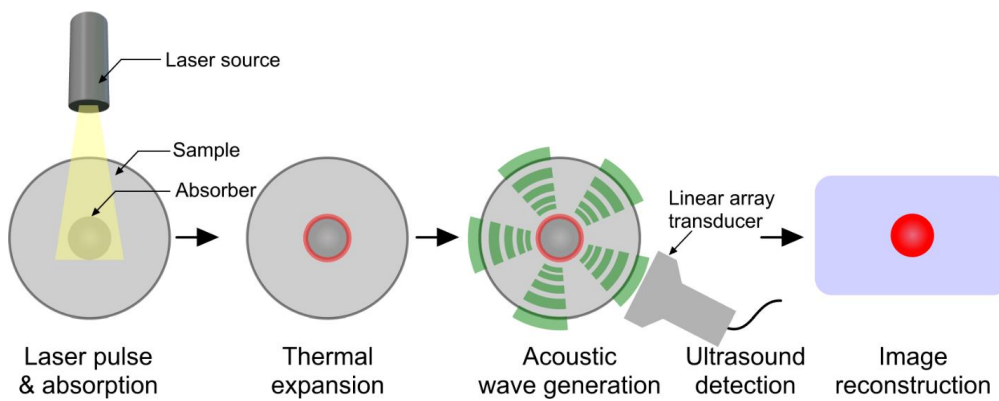

Fig. 2. Principle of PA signal generation, detection, and image reconstruction. Reprinted with permission from Ref. [7]. A high energy and short pulse laser light illuminates the absorber, leading to a transient temperature rise which results in a thermoelastic expansion of the absorber, and acoustic (or PA) wave generation. The signals generated from the waves received by an ultrasound probe are given to a reconstruction algorithm to form a PA image.

Due to strong optical scattering in biological tissues, pure optical imaging modalities have a shallow imaging depth [1, 8-10]. The transport mean free path (i.e., the mean distance after which a photon's direction becomes random) in biological tissues is around 1 mm [3]. Acoustic waves experience far less tissue scattering, thus they propagate a greater distance [11, 12]. Although ultrasound imaging can image deep in biological tissues with a high spatial resolution, its acoustic contrast is incapable of providing certain physiological parameters [13]. In PAI, there is no restriction for photons, thus optical excitation can travel far beyond the diffusion limit and still generate acoustic waves. Sensitivity of PAI in deep tissues is orders of magnitude higher than that of pure optical imaging modalities [13, 14]; the highest penetration depth reported in PAI is ~12 cm [15-17]. PAI is an ideal modality for measuring or monitoring tissue physiological parameters by imaging the concentration of tissue chromophores [18], which are changed during the course of a disease [19]. PAI has been evaluated in preclinical and recently in clinical applications for disease detection and monitoring purposes [20-41]. For instance, it has been used to study human skin abnormalities [37, 42-44], brain disease detection [45, 46], human breast tumor detection [19, 47-50], retina disease diagnosis [51, 52], and atherosclerosis evaluation of vessel walls [30, 53, 54].

PAI can be implemented in tomography or microscopy configurations with variations in system design (more details are provided in [7]). Among all configurations and variations of PAI, linear-array based PAI is one of the most commonly used due to its straightforward setup, easy use, and simple clinical translatability. In addition, since linear-array ultrasound imaging systems have been well-established as clinical tools, slightly modifying them to develop more capable tools, i.e., US/PA imaging systems, is not far from reality.

In recent years, many researchers have used the Vantage Verasonics® system for simultaneous US and PA imaging. Vantage system is a MATLAB-friendly programmable platform that can produce co-registered US and PA images; the PA image represents the optical absorption map of the tissue while the US image represents the tissue acoustic impedance map. Here, we describe the subtle details of US/PA imaging system setup, study the performance parameters of the system, and explain sequencing of the US/PA signal generation and signal amplification as well as the details required for efficient use of the hardware of the system and data processing protocols.

## 2. Vantage system architecture for US/PA sequencing

Vantage hardware is controlled by MATLAB coding. The detailed explanation of the parameters used in the code for data acquisition and processing, image reconstruction, and display are given in Appendix B. The Vantage system architecture for US/PA sequencing is shown in Figure 3A. The structures for US and PA run-time defined in the MATLAB code communicate to *runAcq* through the activation of the MATLAB loader function VSX. This opens a GUI including imaging windows of the reconstructed images from *RunAcq* (US and PA), time gain control, US voltage transmit, and many other parameters that are used for real-time user modification to the imaging sequence. These modifications take effect on the hardware between cycles of hand-offs between VSX and *RunAcq*. After initial activation of VSX, *RunAcq* then communicates all of the run-time transmit and receive structures to the Vantage hardware (silver box) through the PCIe cable. *RunAcq*, receives the raw data, stores it in the RcvBuffer variable, and performs image reconstruction on US and PA data and sends it back to the VSX GUI. The raw data of US and PA are accessible after completion of the imaging sequence [55].

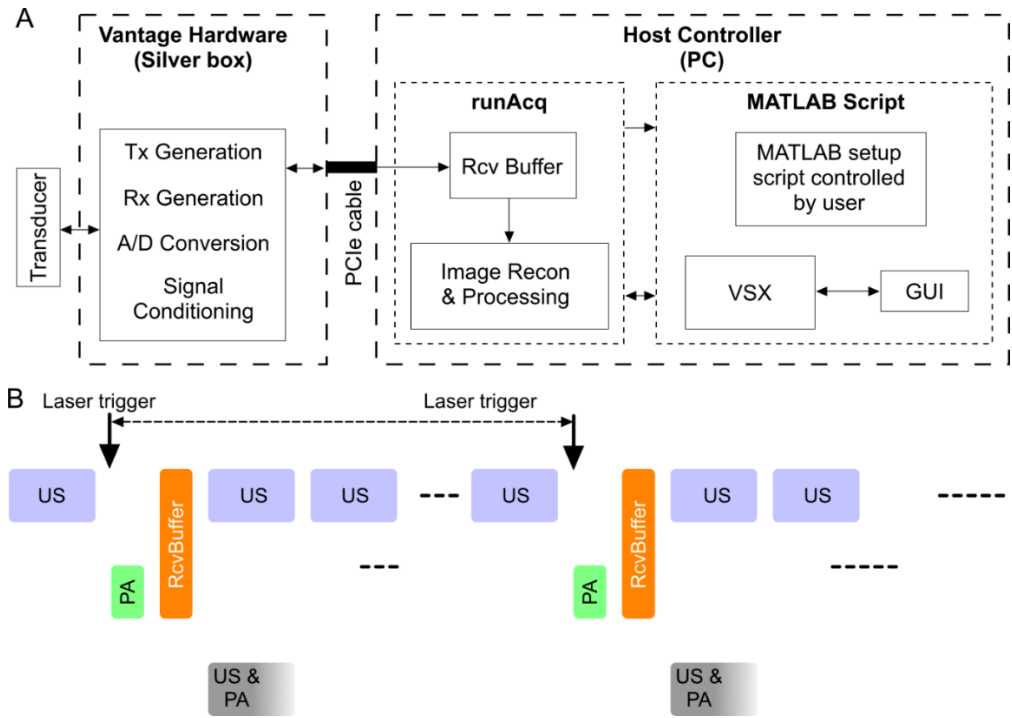

Fig. 3. Vantage system architecture and data acquisition sequencing. **(a)** Vantage system architecture for US/PA sequencing and **(b)** timing diagram of US/PA sequencing in Vantage system. US: Ultrasound, PA: Photoacoustic, Recon.: Reconstruction. RcvBuffer is the predefined buffer where the US and PA data are stored, Acq: acquisition, TX: transmit, RX:

The timing sequence in US and PA data acquisition, illustrated in Figure 3B, is as follows. US acquisition requires a user defined number of steered angles. Between each set of US acquisitions, the system waits for a laser trigger input to begin one PA event. After PA acquisition, US and PA data are transferred to the host controller and stored in the RcvBuffer variable. RunAcq then performs image reconstruction on the averaged data of all of the steered angles and also on the PA data. The reconstructed US and PA images are then displayed in the MATLAB VSX GUI.

## 3. US/PA experimental setup and system characterization

To perform an US/PA experiment, an Nd:YAG laser (PhocusMobil, Opotek Inc., CA, USA) with a repetition rate of 10 Hz and a pulse width of 8 ns was used. The laser uses an optical parametric oscillator (OPO) to tune the wavelength of light between 690 nm to 950 nm. For light delivery, we used a custom fiber bundle (Newport Corporation, Irvine, CA, USA). For data acquisition, the Verasonics Vantage 128 system was used. The specifications of the vantage system are listed in Table 1.

**Table 1. High frequency Vantage 128 system specifications (Verified by Verasonics' technical support)**

| Parameters | Specification |
| --- | --- |
| Channels | 128 Tx / 128 Rx |
| Power requirements | 100V-240V (50-60 Hz) |
| Programmable pulser voltage | 2 to 190 V p-p |
| Time delay resolution | 4.0 ns |
| HIFU capability | Yes |
| A/D resolution | 14 bits |
| Max sampling rate | 62.5MHz |
| Fastest external trigger | 100 KHz |
| Maximum trigger voltage level | 5.5 V |
| Single transmit event | ~ µsec |
| Data pre-amplification | 24 dB |
| Second stage Amplification | 30 dB |
| Noise Figure | 1.5 to 3.0 dB |
| Data transfer to host computer | via 8 lanes PCIe 3.0 sustained data transfer rates up to 6.6 GB/s |

The Vantage system is connected to a host computer using a PCIe express cable. A transducer is connected to the 260 pin Cannon connector mounted on the Vantage system. On the sample end, the transducer was placed and held perpendicularly to the sample. There is a coupling media between the imaging target and the transducer. Water and ultrasound gel are the most common couplant [56]. More details about acoustic couplants can be found in Appendix C.

The Verasonics software package provides a PA MATLAB script, *SetUpL11_4vFlashAngPA.m*, for the 128-element linear array L11-4v probe (see Appendix B). This script can be modified to be used with other probes. Two commonly used ultrasound transducers with Vantage system are ATL Philips L7-4 and Verasonics L22-14v. The specifications of these transducers are listed in Table 2.

**Table 2. Specifications of linear array transducers ATL Philips L7-4, and Verasonics L22-14v**

| Model (Unit) | L7-4 | L22-14v |
|---|---|---|
| Lens correction | 0.8870 | 0.6804 |
| Central frequency (MHz) | 5.2080 | 15.6250 |
| Number of Elements | [4, 7] | [14, 22] |
| Element width (mm) | 128 | 128 |
| Spacing (mm) | 0.2500 | 0.0800 |
| Max high voltage (V) | 0.2980 | 0.1 |
| Focal length (mm) | 50 | 30 |
| Photograph of the device | 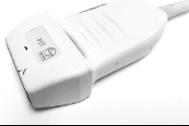 | 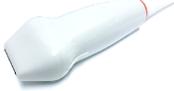 |
| Dimension | (11cm × 6.0cm × 3.0cm) | (9cm × 3.5cm × 2.5cm) |

Table 3 lists the modifications on the original code, written for the L11-4v, needed for the L7-4 and L22-14v. The remainder of the script is compatible as long as the probe is a 128-element linear array.

**Table 3. Modifications to original L11-4 script by Verasonics, for L7-4 and L22-14v probes**

| Parameters | Original Script | Modified Script |
|---|---|---|
| Imaging depth | P(1).startDepth = 2;<br>P(1).endDepth = 192;<br>P(2).startDepth = 0;<br>P(2).endDepth = 128; | P(1).startDepth = 0;<br>P(1).endDepth = 192;<br>P(2).startDepth = 0;<br>P(2).endDepth = 192; |
| Laser parameters | oneway = 0;<br>flash2Qdelay = 200;<br>PA_PRF = 100; | oneway = 1;<br>flash2Qdelay = 188;<br>PA_PRF = 10; |

| Active transducer | *Trans.name = 'L11-4';* | *Trans.name = 'L7-4';\** |
|---|---|---|
| | *Trans.units = 'mm';* | *Trans.units = 'mm';* |
| | *Trans = computeTrans(Trans);* | *Trans = computeTrans(Trans);* |
| | *nElem = Trans.numelements;* | *nElem = Trans.numelements;* |
| | *Trans.maxHighVoltage = 50;* | *Trans.maxHighVoltage = 50;\** |

*Trans.name = 'L22-14v' and Trans.maxHighVoltage = 30 in the case of the L22-14v.

The Vantage system supports linear array transducers from different companies. The list of different transducer arrays that can be used with the Vantage system is given in Table 4. If an unrecognized transducer needs to be used, the transducer attributes must be determined independently and input into the *computTrans.m* script under a new custom transducer case. Key attributes include the central frequency, bandwidth, number of elements, element width, element spacing, element position, and connector pinout arrangement. In addition to unrecognized linear arrays not listed in Table 4, single-element-based arrays, ring arrays, and hemispherical arrays can also be connected to the Cannon connector mounted on Vantage system via a custom-made converter.

**Table 4. List of immediately recognizable transducers for the Vantage system**

| Parameters | Transducer Model Number | |
|---|---|---|
| Philips | L7-4 | C8-4V* |
| | L11-5 | C8-5* |
| | L12-5 38mm | C9-5ICT |
| | L12-5 50mm | P3-2* |
| | CL10-5 | P4-1 |
| | CL15-7 | P4-2 |
| | C4-2 | P5-3* |
| | C5-2 | P6-3 |
| | C7-4* | P7-4 |
| Verasonics | L10-4v | L22-14vX |
| | L11-4v | L22-14vX-LF |
| | L11-5v | L35-16vX |
| | L12-3v | L38-22v |
| | L22-8v | C5-2v |
| | L22-14v | P4-2v |

| | L22-14vLF | |
|---|---|---|
| | GE9LD | GEIC5-9D |
| GE | GEC1-6D | GEL3-12D |
| | GE4CD | GEM5ScD |
| Siemens | L10-5 | |
| Vermon | 4DL7* | |

*These transducers are recognizable by the system, but the system does not have the transducer's attributes which must be manually entered into the computeTrans.m MATLAB file.

To demonstrate the PA signal characteristics in both time and frequency domains, a 2 mm diameter carbon lead phantom in water was imaged at 690 nm. The energy after the fiber bundle was measured as ~15 mJ/cm$^2$. L22-14v transducer array was used in this experiment. The time domain signal obtained from the central transducer (i.e., 64$^{th}$ element), is shown in Figure 4A. In Figure 4B, the magnitude of the single-sided Fast Fourier Transform (FFT) of the time domain signal is shown. In Figure 4A, we also show that the strength of the US waves reflected back from the phantom is greater than that of the photoacoustic [57, 58].

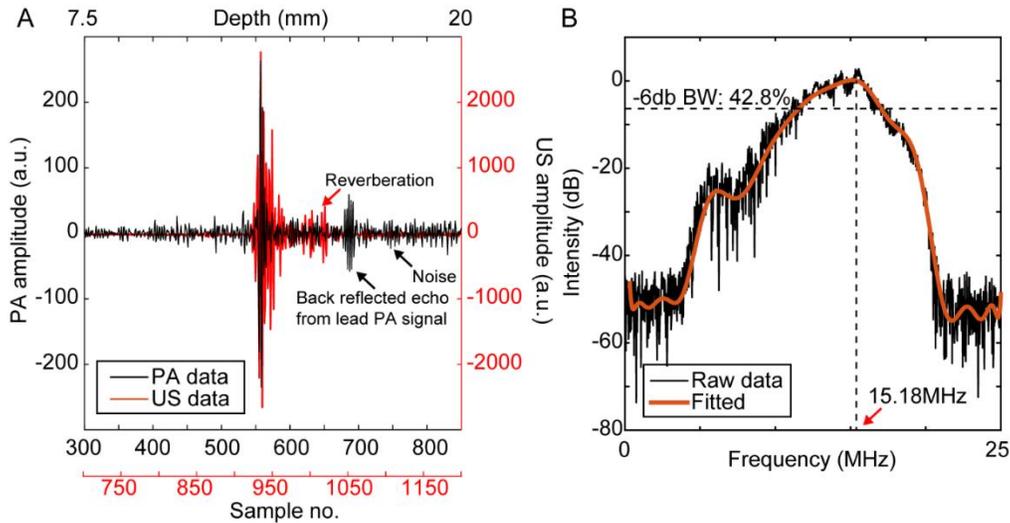

Fig. 4. Photoacoustic signal in (A) time-domain, and (B) frequency-domain, obtained from a 2 mm diameter carbon lead phantom, imaged with L22-14v transducer at 690 nm wavelength. The red signal in A shows the overlaid US signal acquired from the same sample.

We characterized L7-4 and L22-14v for US/PA imaging. We created a resolution phantom of hair of diameter 54 μm in an open top plastic cubic box filled with deionized water (see Figure 5A). The lateral and axial resolutions were quantified by measuring the full width half maximum (FWHM) of the normalized 1D intensity profile across the hair image [59]. The axial and lateral resolutions of the L7-4 were measured at depths 0.5 to 5.0 cm with steps of 0.5 cm (see Figure 5B). The axial and lateral resolutions of the L22-14v were measured at depths 0.2 to 2.0 cm with steps of 0.2 cm (see Figure 5C). In L7-4, the axial resolution remained constant in both US and PA imaging while the lateral resolution worsened with depth, whereas in L22-14v, the axial and lateral resolutions stayed almost constant in both US and PA images. This is potentially due to the focal depth specified by the Vantage system. High frequency transducers have less beam divergence than low frequency transducers, this may have also contributed to worsening the lateral resolution of the L7-4 and constant lateral resolution of the L22-14v [60].

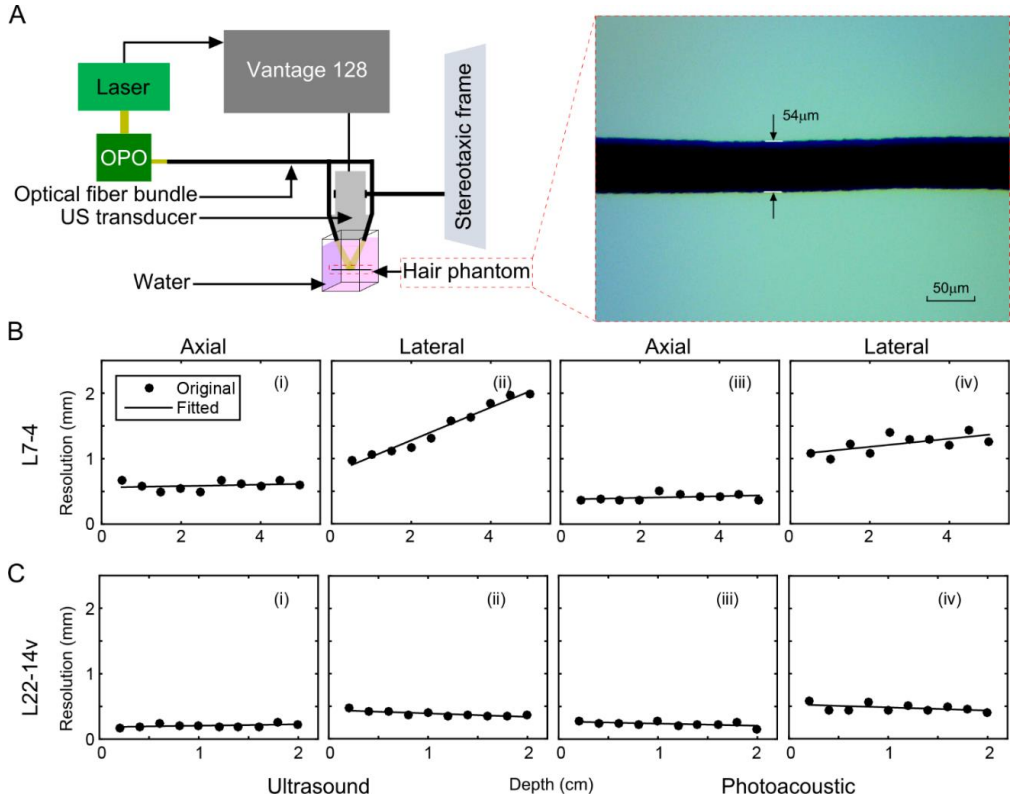

Fig. 5. US/PA resolution study. (A) Schematic of the experimental setup, including a hair phantom photograph captured by a 4× objective on a light microscope (SME-F8BH, Amscope, CA, USA). Resolution study when (B) L7-4 was used, (C) L22-14v was used. (i) US axial resolution versus depth, (ii) US lateral resolution versus depth, (iii) PA axial resolution versus depth, (iv) PA lateral resolution versus depth.

## 4. Technical Considerations

### 4.1 Synchronization Optimization for Efficient Averaging

Photoacoustic imaging relies on careful timing of data acquisition and laser firing to accurately reconstruct photoacoustic depth information. In Q-switched Nd:YAG lasers, the timing involves triggering of the flash lamps to stimulate the emission medium (Nd:YAG) followed by time for optical buildup and subsequent opening of the Q-switch for release of the laser beam. A straightforward triggering method (see Figure 6A (i)) involves using internal flash lamp and Q-switch triggers built into the laser being used and using the flash lamp out port to trigger data acquisition on the Vantage trigger in-1 port. Upon receipt of the flash lamp trigger, the Vantage waits for a user-specified delay time equal to the known optical build-up time for the laser and then begins recording approximately when the laser fires. The major drawback to this triggering method is that the optical build-up time for the laser will "jitter" between tens of nanoseconds and will cause the peak PA signal to fluctuate between a few sample numbers. This will cause issues in data averaging between subsequent frames. We have implemented a more optimal triggering method following the method described in [61] (see Figure 6A (ii)). In this method the flash lamp and Q-switch of the laser are both externally triggered. Using a function generator (ATF20B, Shenzhen Atten Electronics Co., Ltd., Nanshan, China) a trigger of 10Hz, 5Vpp, 2.5V offset, 50% duty cycle was simultaneously sent to the flash lamp-in of the Nd:YAG laser (PhocusMobil, Opotek, CA, USA) and the trigger in-1 of the Vantage system. Within the Vantage MATLAB script, the f*lash2Qdelay* parameter is set to the external Q-switch delay time specified by the laser retailer (290 µs). This allows the Vantage to then trigger the Q-switch by connecting the Vantage's trigger-out to the Q-switch-in on the laser. We have compared the performance of the two triggering methods by imaging a 2mm diameter carbon lead, when L7-4 US probe was used for PA signal detection and the fiber bundle we described earlier was used for illumination at 690 nm. Five sequential photoacoustic frames were recorded and overlaid with both methods (see Figures 6B(i) and 6B(ii)). The results show a greater "jitter" in the first triggering method compared to nearly no "jitter" in the second method. We then averaged the signals acquired with each of these triggering methods. Averaging improved the signal-to-noise ratio (SNR) of the PA signal probably because the signal components being averaged were correlated and the noise components were not. 64$^{th}$ element of the L7-4 probe data was averaged using various numbers of stored frames: 1, 10, and 50, and plotted (see Figure 6B(iii)).

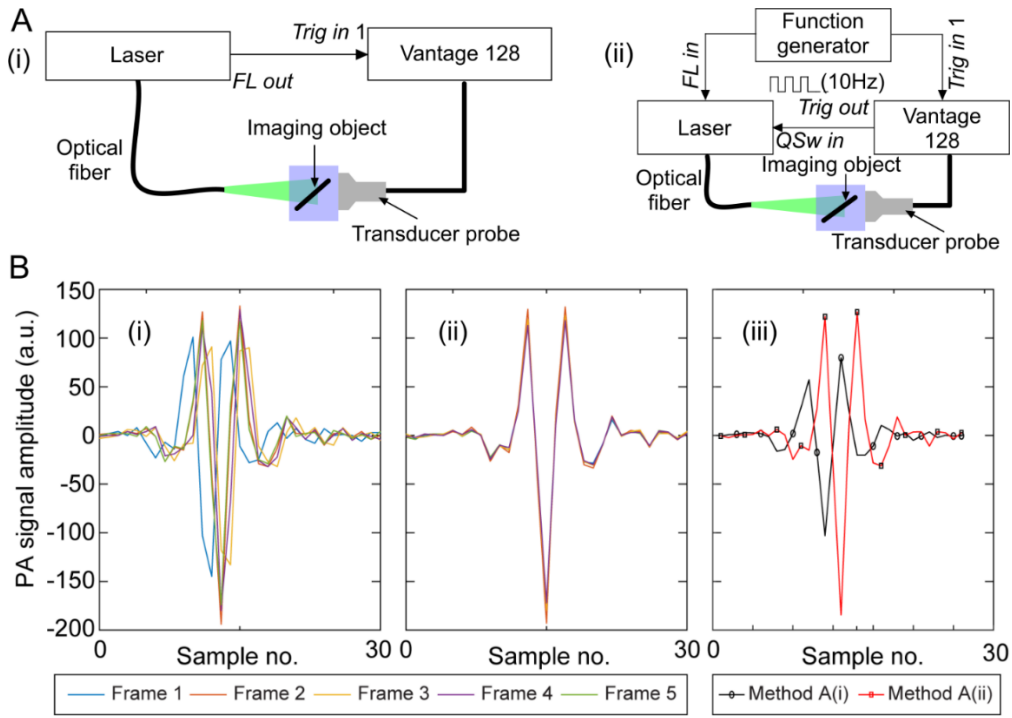

Fig. 6. Photoacoustic signal "jitter" comparison between two triggering methods when a 2 mm carbon lead was imaged. 64$^{th}$ element signals are plotted. (A) (i) Schematic of straightforward triggering method where laser triggers Vantage system, (ii) schematic of the function generator driven triggering method where function generator triggers laser flash lamp and Vantage system, which then triggers laser Q-switch, (B) (i) five sequential PA frames showing significant "jitter" between frames, (ii) five sequential PA frames showing nearly no "jitter" between frames, and (iii) averaged frames using methods described in A(i) and A(ii).

*4.2 Illumination Angle Matters*

Light illumination is one of the main components of a photoacoustic imaging system [62]. The orientation of the illumination of light can influence the amplitude of the PA signal received by the transducer based on how the light energy is deposited on the object [12]. To investigate this matter, we imaged a 2 mm diameter carbon lead in a scattering media, i.e., 2% milk, at 690 nm. We acquired PA data at different illumination angles. The angles were measured as the angle between the transducer and the illumination plane (see the experimental setup in Figure 7A). In these experiments, the transducer plane was held constant (always perpendicular to the sample) and the illumination plane was changed while the illumination spot was at the same location on the surface of the phantom. Figure 7B (i) shows the data collected from the $64^{th}$ element of the L7-4 probe at an angle of $\theta = 40°$ where the PA signal was weakest. Figure 7B (ii) shows the data collected from the $64^{th}$ element of the L7-4 probe at an angle of 48° where the PA signal was strongest. The difference between the amplitudes of the PA signals is probably related to: (i) the light illumination path at each angle which depends on the scattering map of the phantom, (ii) the mechanical property of the tissue as to which directions the thermoelastically generated pressure waves have the maximum amplitude, and (iii) at greater angles, larger area of the target phantom is illuminated, hence, more numbers of point sources are excited and collectively a larger pressure wave is generated. It can be seen that the shape of the two induced PA signals are consistent while the PA signal amplitude is higher at one angle over another. This demonstrates the importance of optimizing illumination in PA imaging.

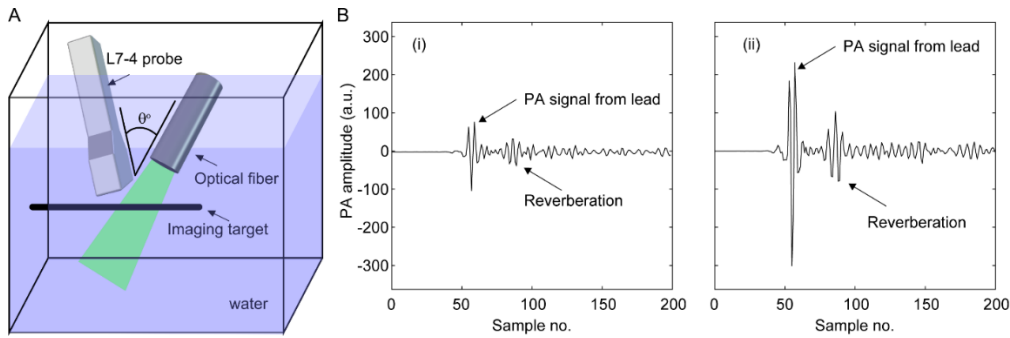

Fig. 7. Impact of illumination angle. (A) Experimental setup of illumination angle investigation. (B) PA signal profile from $64^{th}$ element of L7-4 probe taken from a 2 mm carbon lead phantom with different illumination angles. (i) 40 degrees, (ii) 48 degrees.

*4.3 Improving Reading Accuracy by Upsampling*

The Vantage system automatically samples signals at a rate that is 4 times the central frequency of the transducer being used. This is sufficient to meet the Nyquist limit, however it may be necessary to analyze PA signals at a higher sampling rate. This requires modifications to the script as well as the use of an in-built filter tool to change the spectrum of the bandpass filter on the receive end of the transducer so that the system does not mistake the higher sampled signals as high frequency noise. First, the *Resource.RcvBuffer(1).rowsPerFrame* parameter is multiplied by the factor that the sampling rate will be multiplied. This is to allow the data buffer to hold the increased number of samples. Similarly, the *Receive.decimSampleRate* parameter is multiplied by the same factor. Finally, the parameter *Receive.inputFilter* is modified by entering *filterTool* into the MATLAB command line. This brings up a GUI which outputs the *Receive.inputFilter* value based on the user input parameters. Inside the GUI, *sampleMode* are set to custom with the *decimSampleRate* set to the rate of choice. The bandpass filter central frequency and relative bandwidth are then set to the standard values of the probe being used.

The modifications to the script can be seen in Table 5 for using the L7-4 probe and changing the sampling rate from 20.8MHz to 62.4MHz.

**Table 5. Script modifications to change the sampling rate of L7-4 from 20.8 MHz to 62.4 MHz. The modifications are indicated in red**

| Modification no. | Original | Modified |
|---|---|---|
| 1 | Resource.RcvBuffer(1).rowsPerFrame = 2048*(na + ne); | Resource.RcvBuffer(1).rowsPerFrame = 3*2048*(na + ne); |
| 2 | Receive = repmat(struct('Apod', ones(1,Trans.numelements), ... 'startDepth', P(1).startDepth, ... 'endDepth', P(1).startDepth + wl4sPer128*ceil(maxAcqLngth2D/wl4sPer128), ... 'TGC', 1, ... 'bufnum', 1, ... 'framenum', 1, ... 'acqNum', 1, ... 'sampleMode', 'NS200BW', ... 'mode', 0, ... 'callMediaFunc', 0), 1, (na+ne)*Resource.RcvBuffer(1).numFrames); | Receive = repmat(struct('Apod', ones(1,Trans.numelements), ... 'startDepth', P(1).startDepth, ... 'endDepth', P(1).startDepth + wl4sPer128*ceil(maxAcqLngth2D/wl4sPer128), ... 'TGC', 1, ... 'bufnum', 1, ... 'framenum', 1, ... 'acqNum', 1, ... 'sampleMode', 'NS200BW', ... 'mode', 0, ... 'callMediaFunc', 0,... 'inputFilter', [+0.00000 +0.00122 +0.00400 +0.00687 +0.00677 +0.00378 +0.00323 ... +0.00931 +0.01569 +0.00861 -0.01431 -0.03320 -0.02631 -0.00491 ... -0.01303 -0.07413 -0.14075 -0.12170 +0.02014 +0.20425 +0.28894],... 'decimSampleRate', 20.8333*1), 1, (na+ne)*Resource.RcvBuffer(1).numFrames); |

We imaged a black tape of 18 mm width at 3 sampling rates, 20.8, 41.6, and 62.4MHz. The tape was imaged in a water bath with the L7-4 probe at the illumination wavelength of 690 nm. The results for these sampling rates are given in Figure 8A, B, and C, respectively. To demonstrate the improvement in the reading of the signals with increasing the sampling frequency, we calculated the difference between the top peaks and also the bottom peaks in the signal. The values are indicated in each figure. One drawback of using higher sampling frequency is slower available frame rate, therefore, an alternative solution to increase the number of samples without increasing the sampling frequency is interpolation. In Figure 8D, we showed the interpolated signal (3 times up sampling) using the 'spline' algorithm [63]. We also provided the location (i.e., index) of the highest peak value normalized by respective sample numbers ( $x_{nor} = x(y_{max})/\text{sample no.}$ ) and the ratio of peak values (both positive and negatives).

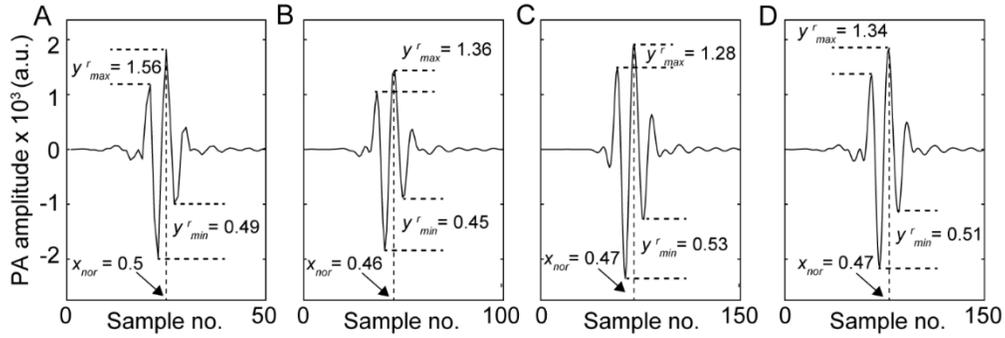

Fig. 8. PA signal profile from 64th element of L7-4 probe taken from a black tape phantom of 18mm width with varying sampling rates: (A) 20.8 MHz, (B) 41.6 MHz, (C) 62.4 MHz, and (D) interpolated signal using 'spline' algorithm. Here, $x_{nor} = x(y_{max})/\text{sample no.}$ and $y^r_x = $ Ratio of $y$ values, where $y = $ positive or negative peaks.

## 4.4 Simultaneous Real-time Visualization of US and PA Images

The default MATLAB script provided by Verasonics® utilizes a single imaging window for both US and PA image display, with the user being able to modify parameters to view one or the other. With real-time imaging, it is more important to simultaneously view US and PA images side by side. The following describes how to modify the original example script to simultaneously display US and PA images in real-time. With a single imaging window, all parameters for the *Resource* structure have arguments of 1. Adding a second window requires duplication of all of the *Resource.DisplayWindow* parameters with the argument of 2. Specifically, the variables *Title, pdelta, Position, ReferencePt, numFrames, AxesUnits, Colormap,* and *splitPalette* are copied from the already existing *Resource.DisplayWindow(1)* structure with the argument changed to 2. To differentiate between US in window 1 and PA in window 2, we set *Resource.DisplayWindow(1).Colormap* to gray(256) and *Resource.DisplayWindow(2).Colormap* to hot. The other change necessary to keep both windows real-time is modification of the *Process(n).Parameters* structure where *n* is 1 for US and 2 for PA. In the original structure, *mappingMethod* is set to *lowerHalf* for US and *upperHalf* for PA, but both should now be set to *full.* Also, *displayWindow* was originally set to 1 in both processes and now it should be 1 for US and 2 for PA. Modifications to the original script to simultaneously display US and PA images, in real-time are given in Appendix D (the modifications are indicated in red).

## 4.5 US Image Quality Improvement

Vantage offers beam steering as a method of improving US image quality. This technique modifies transmit delay times in the beam apodization to steer the beam at different directions into the imaging field. This provides received echoes from different angles of reflection off of the imaging targets. The received echoes from multiple steered angles are then averaged to form an improved image. The variable *na* in the MATLAB script can be set to any positive integer to determine the number of steered angles. The transmit waveforms will be emitted at angles $\frac{n\pi}{N+1}$, where *N* is equal to the variable *na*. *n* iterates with each transmit angle from 1 to *N*, where $\frac{\pi}{2}$ is normal to the transducer elements. Upon definition of *na*, these angles and corresponding time delays in the apodization for each element are calculated, and the time occurs between each US steered angle is determined. Taking this time into account with the pulse repetition rate of the laser, the number of the angles occur between each laser pulse is determined. Further, increasing the number of steered angles and the number of stored frames in each buffer are balanced to avoid slowdown of the system due to memory build up, which can potentially freeze MATLAB if the system RAM is overflowed. To investigate the effect of the number of steered US transmit angles on the image quality, we imaged a phantom, with 2 mm, 0.9 mm, and 0.2 mm diameter carbon leads in water with varying numbers of steered angles (i.e., 8, 16, 32, 64, 128), and quantified the contrast to noise ratio (CNR) of resultant images following the method explained in [12]. Figure 9A demonstrates the experimental setup. Figures 9(B-F) show the US image for 8, 16, 32, 64, and 128 steered angles, respectively. Figure 9G shows the improvement of CNR with greater number of steered angles.

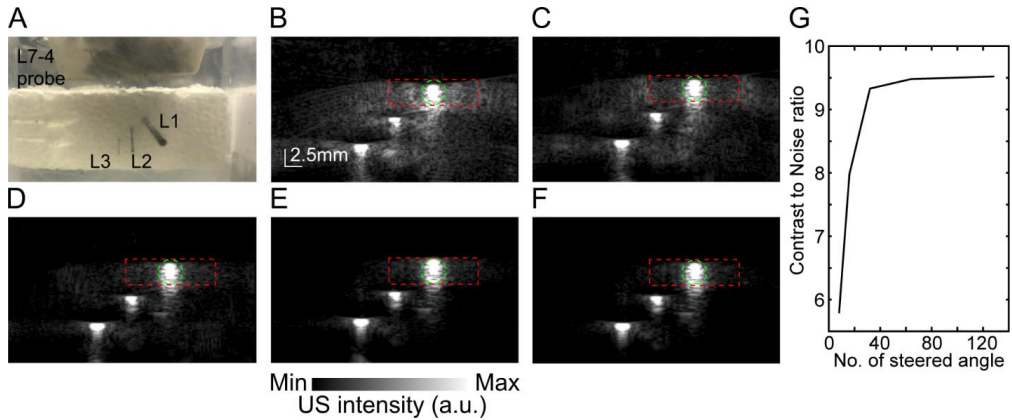

Fig. 9. US image quality improvement. (A) Experimental setup showing an ultrasound probe and a 3-carbon lead phantom. (L1) 2 mm diameter carbon lead, (L2) 0.9 mm diameter carbon lead, (L3) 0.2 mm diameter carbon lead, (B-F) US images of the 3 carbon lead phantom for 8, 16, 32, 64, and 128 steered angles, respectively. Green dashed circle encloses object pixels and red dashed box encloses both object and background pixels, (G) CNR versus number of steered angles

*4.6 Photoacoustic Signal Amplification*

The PA pressure waves received by the linear array transducer are weak (i.e., their SNR is low) [64, 65]. In cases where the imaging media is highly scattering, only weak pressure waves reach the transducer and produce low level voltage signals. These low level signals may not be accurately digitized using the default Analog Front End (AFE) settings of the Vantage data acquisition (DAQ) system. The AFE gain can be increased by up to 12 dB by modifying the settings in the *RcvProfile* structure, however in some cases, additional gain may be required. To increase the amplitude of the received signal, we used a 128-channel amplifier (AMP128-18, Photosound, Houston, TX, USA). The amplifier has 40 dB gain with -6 dB cut off frequencies of 25 kHz and 35 MHz. The amplifier was attached to the 260 pin Cannon connector on Vantage system (see Figure 10). The US probe was then plugged into the front end of the amplifier. In the MATLAB script, i.e., *computeTrans.m*, a custom transducer ID was added. Details of the ultrasound transducer were copied to the custom ID so that the amplifier can be recognized. This is similar to the addition of any unrecognized transducer to the system, only the transducer attributes (e.g., central frequency, bandwidth, number of elements, element width, element spacing, and element position) are determined independently and input into the *computTrans.m* script under a new custom transducer case. To allow for identical wiring path lengths through the amplifier from the transducer end to Vantage system end, the pins were not connected one to one and their organization were corrected in the MATLAB script in the *Trans.Connector* variable. Since the amplifier was programmed to multiply any signal going in or out, to avoid causing damage to the transducer elements during US transmit, the transmit beam was disabled by setting *TX.Apodization* to zeros. This allowed the Vantage system to only work in the receiving mode with amplification. To demonstrate the utility of the amplifier, a two-wire phantom (each with 600 µm diameter) was imaged with and without the amplifier at varying concentrations of intralipid, i.e., 0%, 25%, 50% (Sigma Aldrich, USA). The experimental setup is shown in Figures 10A and B. The setup consists of a 50 Hz, 532 nm Nd:YAG laser (NL231-50-SH, EKSPLA, Vilnius, Lithuania) with 7 ns pulses, and an L7-4 US transducer. The laser illuminated the 2-wire phantom through a large fiber optic bundle (1 cm diameter, FO Lightguide, Edmund Optics, NJ, USA). We observed an SNR improvement in addition to the signal peak increase in all concentrations of intralipid (see Figures 10B, C, and D). SNR improvement was probably due to the high pre-amplifier input impedance which shifted the transducer noise spectrum to low frequencies, which was then filtered out using a high-pass filter [66].

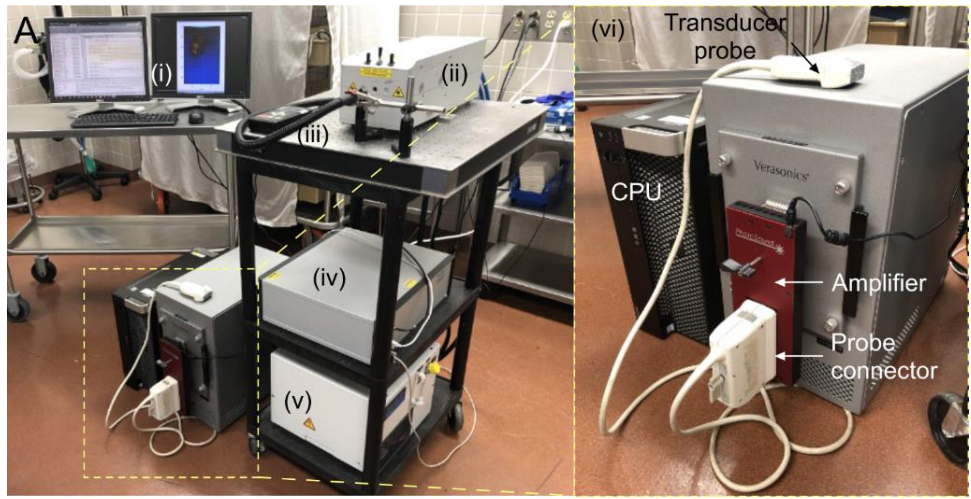

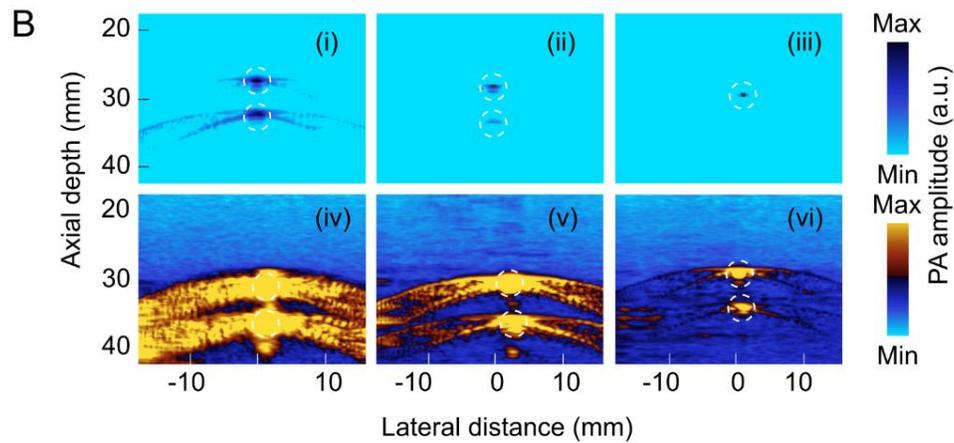

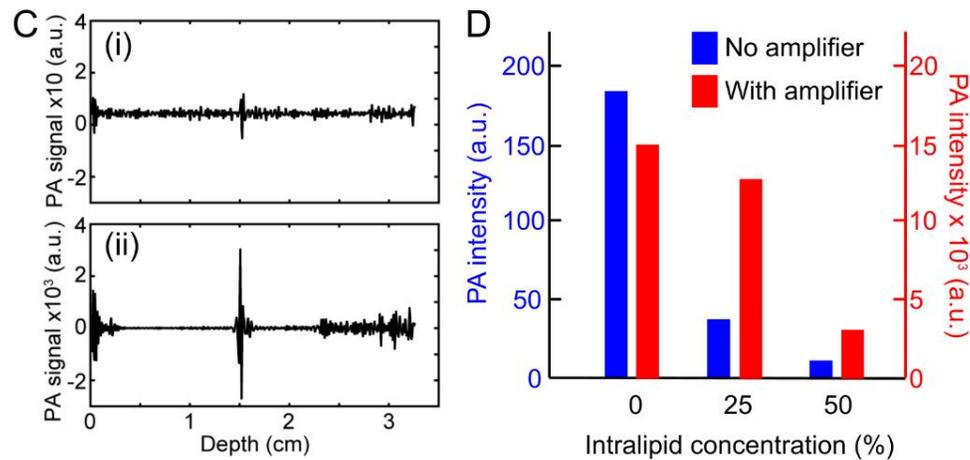

Fig. 10. PA signal amplification. (A) System setup to test the Photosound AMP128-18 amplifier: (i) display monitor, (ii) Nd:YAG laser head, (iii) fiber optic bundle, (iv) laser power supply, (v) laser chiller. (B) Wire imaging results: wire phantom suspended in 0% intralipid imaged (i) without and (iv) with amplifier, wire phantom suspended in 25% intralipid imaged (ii) without and (v) with amplifier, and wire phantom suspended in 50% intralipid image (iii) without and (vi) with amplifier. (C) Raw data from the 64[th] channel of the linear array when imaging a single wire in 50% intralipid concentration to demonstrate SNR improvement from (i) without amplifier to (ii) with amplifier, (D) Bar chart demonstrating signal amplitude increase corresponding to phantom imaging in B with and without amplifier, at different concentrations of intralipid.

## 5. Fluence Compensation

The initial PA pressure is proportional to the product of local optical absorption coefficient and the optical fluence deposited at the same location. However, the optical fluence could vary significantly in living tissue which necessitates the use of optical fluence compensation for quantitative PA imaging [67]. If the optical parameters of the tissue, e.g., absorption coefficient, scattering coefficient, and anisotropy factor, are known, the optical fluence can be calculated by solving the photon diffusion equation using Monte Carlo (MC) or finite element method [68, 69]. With the known optical fluence distribution, a fluence-compensated absorption map can be calculated via dividing the PA reconstructed image by its corresponding fluence map.

We evaluated the fluence compensation algorithm on a phantom made of a gelatin, intralipid, and ink mixture (see Figure 11A and B). We used 0.5 mm diameter carbon lead in three layers of gelatin phantom with thicknesses of 5, 8, and 7 mm, respectively. We added intralipid (Sigma Aldrich, USA) as scatterers and used ink to represent absorption. The concentration of the intralipid solutions were 4%, 1%, and 6%, and those of ink were 0.1%, 0.4%, and 0.2%, respectively; these values were chosen to represent various biological tissues. Using Mie calculator [70], the scattering coefficient, absorption coefficient, and anisotropy factor of the layers were calculated at 532 nm. The scattering coefficients were 15.4 cm-1, 3.1 cm-1 and 22.3 cm-1, respectively. The absorption coefficients were 0.3 cm-1, 0.9 cm-1, and 0.4 cm-1, respectively. The absorption coefficient of water, i.e., 0.11 cm-1, was added to these numbers. We used another layer of pure gelatin with the thickness of 5 mm. The scattering coefficient of both pure gelatin and US gel were considered 0.05 cm-1. The absorption coefficient of US gel (3mm layer) was considered 0.11 cm-1. The experimental protocol was as follows: (1) we imaged the phantom using the L7-4 probe and generated an US image (see Figure 11C) and a PA image (see Figure 11D); (2) we then segmented the US image using the segmentation method given in [71]; (3) the abovementioned optical properties were assigned to each segment, creating the phantom optical model; (4) fluence map (see Figure 11E) of the phantom optical model was generated by an MC simulation (using MCX software [72]); (5) finally, the PA image was divided by the MC simulated image. The resultant image was a fluence-compensated PA image (see Figure 11F). The PA intensity variation between the peaks (at the location of imaging targets) in the fluence-compensated PA image was less than 5%. In this experiment we assumed that the field of view of each element of the transducer is cylindrical.

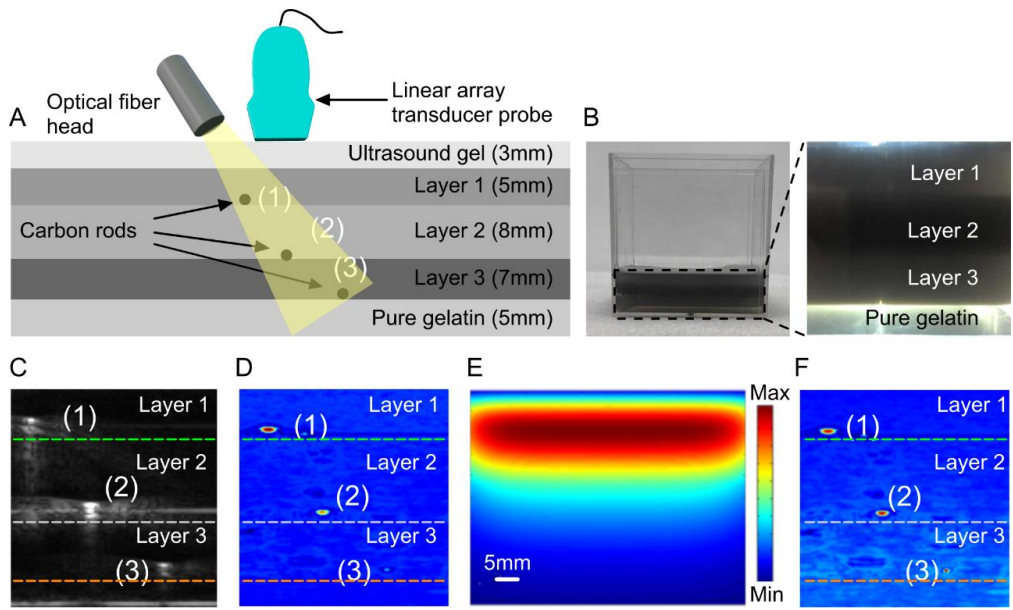

Fig. 11. Fluence Compensation algorithm validation. (A) Schematic of a phantom with 3 layers of gelatin phantom mixed with intralipid/ink solution and 3 carbon lead imaging targets, (B) photograph of the phantom made in a cube box and magnified to show the layers, (C) US image of the phantom, (D) PA image before fluence compensation, (E) Monte Carlo simulation of light propagation in the phantom, (F) PA image after fluence compensation.

## 6. PA Image Reconstruction

After detecting PA signals by the linear array, the channel data are given to an image reconstruction algorithm and an optical absorption distribution map, i.e., PA image, is reconstructed [73-77]. Several image reconstruction algorithms have been studied for linear-array PA imaging [78-81]. The most basic algorithm is delay-and-sum (DAS). DAS follows the dynamic focusing protocol, in which the focus is adjusted for each pixel of the imaging target; the detected signals are delayed proportional to the distance between the focal point and the position of the element in the imaging array. Finally, the delayed signals are summed up, and an image is formed. As a result of this summation, the on-axis signals will overlap on each other while off-axis signals will be suppressed. In the remaining of this section, we used two signal quality metrics to compare the performance of different image reconstruction algorithms: (i) resolution, which is defined as the full width half maximum (FWHM) of the lateral profile of the image of a point source, and (ii) sidelobe, defined as the power of the first lobe next to the main lobe. Although DAS is simple to implement and fast, it treats all the detected signals the same way which results in low resolution images with high sidelobes; this is mainly due to the constructive overlap of the off-axis signals [82]. Delay-Multiply-and-Sum (DMAS) can generate images with a finer resolution and lower sidelobes compared to DAS [83]. Double stage DMAS (DS-DMAS) algorithm offers finer resolution, and lower sidelobes [80, 82, 84] than DMAS; DS-DMAS uses two stages of correlation process to suppress off-axis signals. Minimum variance (MV) significantly improves the image resolution compared to DAS, DMAS and DS-DMAS, however the produced images have high sidelobes; this problem has been addressed in MV-DMAS [78, 85]. The Eigen-space version of MV-DMAS, i.e., EIMV-DMAS, provides similar resolution to MV-DMAS with lower sidelobes [79]. Methods that use MV algorithm, are sensitive to the quality of the received data; this issue limits the application of MV-based algorithms for PA image reconstruction, because PA signals are usually weak. Among the image reconstruction algorithms described above (see Table 6), DS-DMAS is the most suitable one for linear-array based PA image reconstruction, because it reconstructs images with fine resolution and low sidelobes, and has a low computational complexity.

Table 6. Performance comparison between image reconstruction algorithms

| Algorithm | Resolution | Sidelobes | Complexity | Reference |
|---|---|---|---|---|
| **DAS** | Low | High | Low | [82] |
| **MV** | High | High | High | [85, 86] |
| **DMAS** | Low | Medium | Medium | [83, 87] |
| **DS-DMAS** | Medium | Low | Medium | [80, 82, 84] |
| **MV-DMAS** | High | High | High | [78, 85] |
| **EIMV-DMAS** | High | Low | High | [79] |

Four wire phantoms were prepared for this study (Figure 12A). Transducers were securely held by a clamp which was attached to an x-y translation stage. The phantom container was fixed to the optical table and filled with distilled water. Each phantom was imaged 50 times. Figures 12(B-E) demonstrate images of the four wire phantoms taken with the L7-4 probe where (i) is the US image, (ii) is the PA image reconstructed in the Vantage system, and (iii) is the PA image reconstructed with DS-DMAS. Figures 12(F-I) demonstrate images of the four wire phantom taken with the L22-14v probe where (i) is the US image, (ii) is the PA image reconstructed in the Vantage system, and (iii) is the PA image reconstructed with DS-DMAS.

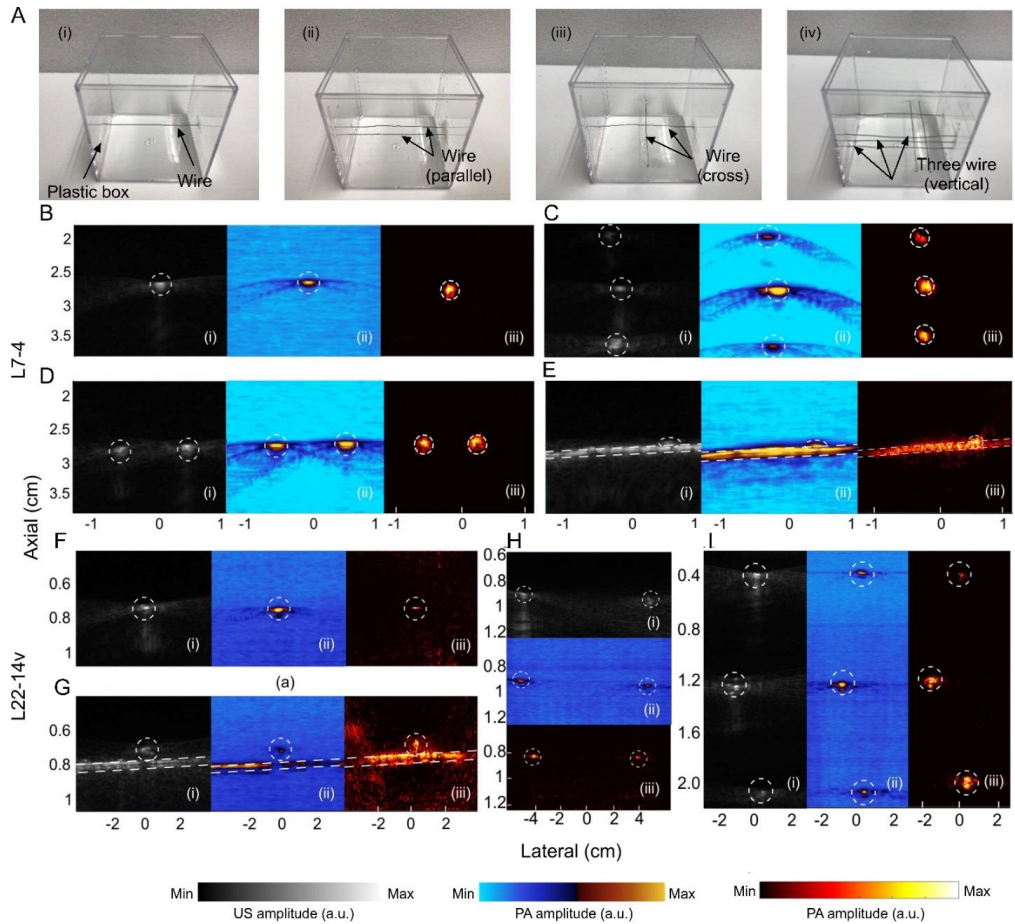

Fig. 12. Comparison of the performance of DS-DMAS to Vantage PA reconstruction method. (A) Wire phantoms for resolution study, (i) one-wire, (ii) two-wire, (iii) two-wire cross, and (iv) three-wire. (B-I) US and PA images produced by L7-4 and L22-14v probes where the image reconstruction is: (i) Vantage default US, (ii) Vantage default PA, and (iii) photoacoustic DS-DMAS. L7-4 probe: (B) single wire cross section, (C) 3-wire cross section, (D) 2-wire cross section, and (E) 2-wire cross section. L22-14v probe: (F) single wire cross section, (G) 2-wire cross section, (H) 2-wire cross section, and (I) 3-wire cross section. Images of the targets are shown in white dotted circles or straight lines.

## 7. Conclusion

Photoacoustic imaging is a complement to the already established US imaging technique and may significantly increase its scope of application in diagnostic imaging and therapeutic monitoring. Combining with commercial medical US systems, the development of PAI can be accelerated by taking advantage of US image reconstruction and processing. Among different configurations, linear-array PAI is becoming popular, mainly because linear-array US transducers can easily be manufactured, hence the production cost is lower as compared to custom-made curved or ring arrays [88, 89]. Moreover, these transducers are commonly used in clinical applications, implying the PAI system built with it will have high clinical translatability.

We discussed technical considerations for US/PA imaging implementation using the Vantage research system. With the information we presented in the body of this technical note and the four appendices, we described most of the experimental considerations one should know when working with the Vantage system for PAI tests. Although the Verasonics Vantage System has many capabilities and advantages, it has some limitations. The transducer connector limits which transducers can directly connect to the system. Verasonics offers some connector adapters to alleviate this problem, but it can make switching between transducers cumbersome. Another limitation is that the signal pre-amplification occurs far (~1-2 meters) from the probe which increases the noise level for weak PA signals. Further, the maximum sampling rate of the system is 62.5MHz which may limit the use of high frequency transducers.

## Appendix A. Absorption Spectrum of Black Tape and Carbon Pencil Lead

Measuring the absorption spectrum of the imaging target, i.e., amplitude of the optical energy versus wavelength, is essential for the analysis of results in PA experiments. Here we explain a method of how to extract the absorption spectra of two example imaging targets, vinyl black tape and carbon pencil lead. The internal energy readings are not proportional to the energy deposited on the sample due to dispersion, i.e., variable attenuation of wavelengths in the optical fiber. Using the PhocusMobil (Opotek, CA, USA) laser, that has a built-in energy meter, we first obtained the *energy ratio* by using a separate energy meter (QE25LP-S-MB-QED-D0, Gentec-EO, Quebec City, Canada) to record the energy after the illumination fiber and dividing it by the internal energy reading inside the PhocusMobil. This yields the relation between the internal Opotek energy reading to the energy deposited on the sample. We imaged a 2 mm carbon lead phantom and a black vinyl tape phantom, and simultaneously recorded the internal laser energies. The PA data was compensated by dividing the raw PA signals by the product of the energies recorded with each phantom and the *energy ratio*. The absorption spectrum of the black tape and carbon pencil lead after performing a fourth order polynomial fit are shown in Figure 13A and B, respectively.

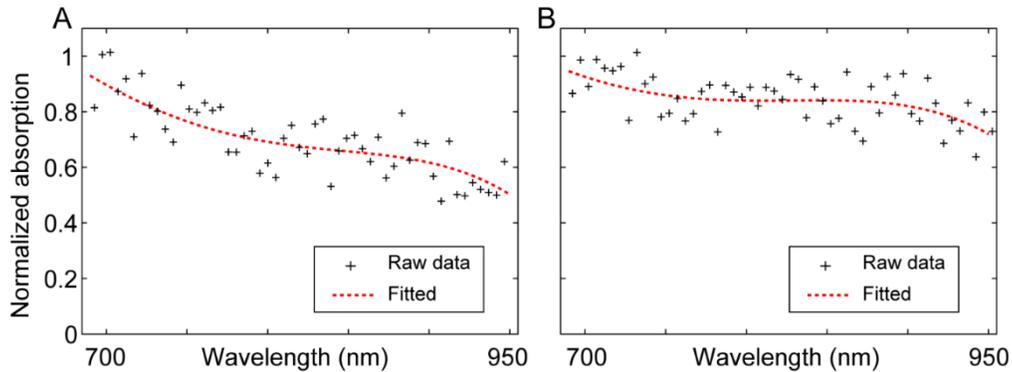

Fig. 13. Measurement of absorption spectrum of two commonly used imaging targets: (A) black tape, and (B) carbon pencil lead.

## Appendix B. MATLAB Codes for Data Acquisition, Processing, Image Reconstruction

MATLAB code for the Vantage system for real-time data acquisition, data processing, image reconstruction, and display in order of variable appearance in the script is provided as we show in S1 (Ref. [90]). Explanation of the code is below:

Start and end depth of imaging are set for US and PA imaging with P(1).startDepth and P(1).endDepth and P(2).startDepth and P(2).endDepth, respectively. na sets the number of steered US angles that are transmitted for a single US frame. The default number is 7. oneway, sets whether or not the system runs in simulation mode: 0 for simulation mode is off or 1 for simulation mode is on. Simulation mode ignores hardware and mimics transducer elements to create virtual signals and test if the MATLAB loader program VSX is communicating correctly with RunAcq. RunAcq communicates receive and transmit commands to the ultrasound hardware. Flash2Qdelay is the time between trigger input and start of acquisition in microseconds and must equal the time delay between the flash lamp output and Q-switch output from the laser. We used an oscilloscope to find the exact delay time and set the time value to the flash2Qdelay variable (see Table 3 for the values we used for our system). PA_PRF is set to the pulse repetition rate of the laser. The next section of the code involves setting up system parameters, such as data buffers, and the transducer array specifications. For Vantage 128 system and a 128-element linear array transducer, Resource.Parameters.numTransmit and Resource.Parameter.numRcvChannels are set to 128. For the two linear array transducers we used, Trans.name is set to either 'L7-4' or 'L22-14v'. ComputeTrans(Trans) populates all the attributes of the specified transducer. PData defines the pixel grid to be manipulated by the image reconstruction software. PData(n).PDelta defines the spacing between pixels in all dimensions, PData(n).Size defines rows, columns, and sections of the data, and PData(n).Origin defines the x,y,z coordinates of the left corner of the reconstructed image, where n is 1 for the US image and n is 2 for the PA image. RcvBuffer, InterBuffer and ImageBuffer are used to store the US and PA raw data and reconstructed images. Transmit waveform (TW), transmit (TX), and transmit power controller (TPC) are transmit objects. TW structure array defines specification of a transmit waveform (type, frequency, duty cycle, duration, and polarity), where TW(1) is for US and TW(2) for PA. The TX structure array defines the specification of the beam characteristics of each transmit action, including which transmitters are active in the aperture (apodization), and the delay time before transmit for each active transmitter. With 7 US transmits, TX(1:7) is defined for US transmit events. For PA, only one TX structure is needed, so we define TX(8) for this event. In TX(8),all transmitters are turned off for the receive-only beamformer. TPC(1) for US and TPC(2) for PA are defined, where TPC sets the transmit power level for each specific transmit event. Receive objects are defined next and populate all the characteristics of the receive phase of an acquisition event. The Transmit and Receive periods both start with the data acquisition. Next, the time-gain-control (TGC) object defines the time-gain-compensation curve for the receive portion of the acquisition event. To define a TGC waveform, the user specifies the TGC.CntrlPts array and TGC.rangeMax. TGC.Waveform is then synthesized and applied to the received data. The next section describes the reconstruction protocol. Recon structure provides the general attributes of the reconstruction, including the source and destination buffers to use. Recon.senscutoff is a value from 0.0 to 1.0 that sets the threshold of sensitivity below which an element is excluded from the reconstruction summation for a given pixel. Recon.pdatanum(n) specifies the number of the PData structure that defines the pixel locations of the reconstructed data, where n is 1 for the US image or is 2 for the PA image. Recon.RcvBufFrame is an optional attribute that when provided, overrides the frame number specified by the ReconInfo structures. Setting Recon.RcvBufFrame to -1 allows the last acquisition frame transferred into the RcvBuffer to be used in the reconstruction and be displayed for real-time imaging. Recon.IntbufDest and Recon.ImgBufDest specify the destination buffer and frame that will receive the reconstructed output, respectively. Recon.RINums is a row vector that specifies the ReconInfo structure indices associated with the most recent reconstruction. For each Recon, there is an associated set of ReconInfo objects which contain information on how to perform the reconstruction. This information includes which data in the data buffer to reconstruct along with where within PData. Further, ReconInfo chooses which type of reconstruction is performed between replace, add, or multiply intensity, where each successive reconstructed frame replaces the previous, is added to the previous, or is multiplied by the previous, respectively. Replace intensity is chosen for our purpose. Process objects are used to describe the type of processing to apply to the acquired data. After defining the sequence control objects, event objects, and some graphical user interface (GUI) controls, the script is complete for data acquisition and displaying US and PA data.

**Appendix C. Optical Absorption of acoustic coupling medium (Water)**

In a PA imaging experiment, there are ultrasonic transducers and an acoustic coupling layer between the transducers and the imaging target [91]. Propagation of generated ultrasound waves from the imaging target is least attenuated when received by the ultrasound sensor through a coupling agent that exhibits minimal acoustic impedance mismatch with the imaging target. The coupling agent helps minimize degradation and signal loss [92]. Acoustic couplants can be characterized as liquid, gel, semi-dry, and dry. Liquids and gels generally have lower acoustic impedance than dry couplant materials. More details about different acoustic couplants can be found in the literature. Here we describe details on the most widely used acoustic couplant, water. Water has a low acoustic attenuation and impedance (1.5 MPa.s/m) which makes it suitable as an acoustic coupling material. In terms of water optical properties, it absorbs over a wide range of the electromagnetic radiation spectrum with rotational transitions and intermolecular vibrations responsible for absorption in the microwave ($\approx$ 1 mm - 10 cm wavelength) and far-infrared ($\approx$ 10 μm - 1 mm), intramolecular vibrational transitions responsible for absorption in the infrared ($\approx$ 200nm- 10 µm), and electronic transitions responsible for absorption in the ultraviolet region (< 200 nm) [93-95]. In regular water ($H_2O$), the first large absorption band occurs at around 980 nm [93, 96, 97], followed by another band at ~1450 nm. The absorption spectrum of heavy water ($D_2O$) is different from that of $H_2O$, mainly due to the heavier deuterium nucleus; absorption peaks occur at around 1000 nm, 1330 nm, and 1600 nm [98, 99]. The spectral features of water absorption also depend upon the temperature [93, 98, 100]; as temperature decreases, the fraction of hydrogen-bound water molecules is increased, causing absorption peaks to reduce in intensity, broaden in bandwidth, and shift to lower energy [96]. The optical absorption spectrum of regular water and heavy water at various temperatures are shown in Figure 14.

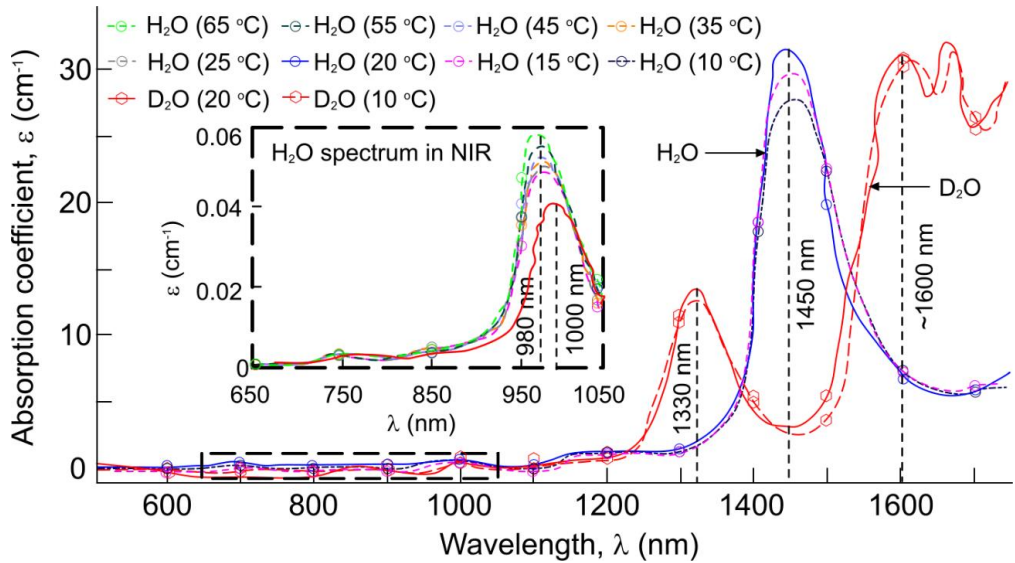

Fig. 14. Optical absorption spectrum of regular water ($H_2O$) and heavy water ($D_2O$) at various temperatures in NIR region. Data for this figure has been extracted from [93, 98] and reproduced.

**Appendix D. MATLAB Code for Simultaneous Diplay of US and PA Image**

The modifications to the original script to simultaneously display US and PA images, in real-time (more details are described in Section 4.4.), as we show in S2 (Ref. [101]).